# Hypermultiplexed Integrated-Photonics-based Tensor Optical Processor

Shaoyuan Ou[1], Kaiwen Xue[1], Lian Zhou[1], Chun-ho Lee[1], Alexander Sludds[2], Ryan Hamerly[2, 3], Ke Zhang[4], Hanke Feng[4], Reshma Kopparapu[1], Eric Zhong[1], Cheng Wang[4], Dirk Englund[2], Mengjie Yu[1,5], Zaijun Chen[1,5,6]*

[1]Ming Hsieh Department of Electrical and Computer Engineering, University of Southern California, Los Angeles, CA 90089, USA
[2]Research Laboratory of Electronics, MIT, Cambridge, MA 02139, USA
[3]PHI Laboratories, NTT Research Inc., 940 Stewart Drive, Sunnyvale, CA 94085, USA
[4]Department of Electrical Engineering & State Key Laboratory of Terahertz and Millimeter Waves, City University of Hong Kong, Kowloon, Hong Kong, China
[5] School of Advanced Computing, University of Southern California, Los Angeles, CA 90089, USA
[6] Opticore Inc., Los Angeles, CA 90007

*Email: zaijunch@usc.edu, zaijun@berkeley.edu

**Abstract:** The escalating data volume and complexity resulting from the rapid expansion of artificial intelligence (AI), internet of things (IoT) and 5G/6G mobile networks is creating an urgent need for energy-efficient, scalable computing hardware. Here we demonstrate a hypermultiplexed integrated-photonics-based tensor optical processor (HITOP) that can perform trillions of operations per second (TOPS) at the energy efficiency of 40 TOPS/W. Space-time-wavelength three-dimensional (3D) optical parallelism enables $O(N^2)$ operations per clock-cycle using $O(N)$ modulator devices. The system is built with wafer-fabricated III/V micron-scale lasers and high-speed thin-film Lithium-Niobate electro-optics for encoding at 10s femtojoule/symbol. Lasing threshold incorporates analog inline rectifier (ReLu) nonlinearity for low-latency activation. The system scalability is verified with machine learning models of 405,000 parameters. A combination of high clockrates, energy-efficient processing and programmability unlocks the potential of light for large-scale AI accelerators in applications ranging from training of large AI models to real-time decision making in edge deployment.

**Teaser:** Space-time-wavelength computing with micron-lasers on thin-film lithium niobate enables scalable optical processing at femtojoule per operation.

## 1. Introduction

Tensor processors, with their unique prowess in data-intensive algorithms, have become a major computing building block in modern high-performance computing (HPC) and artificial intelligence (AI), leading science and technology innovations in deep learning [1] for language processing [2], image analysis [3] and automation [4], NP-hard optimization [5] for scheduling, route planning, and resource allocation, and iterative solvers [6] for multiphysics simulation [7] and cryptocurrency [8] (Fig. 1a). In particular, recent success in large language models (LLMs) [9] has made AI available in our daily lives, improving our work efficiency and convenience [10]. Typically, LLMs with up to a trillion parameters (GPT-4, as of 2023 [11]) are trained with trillions of input tokens (texts or codes), each computes with the model parameters leading to demanding requirements in computing hardware, energy cost, and training times (up to several months [11]). The computing power requirement has become the fundamental bottleneck in the further development and deployment of AI models.

Due to the massive memory interface requirements in tensor processing [12], traditional Von-Neumann architectures are inefficient in performing these tasks. New computing paradigms (Fig. 1b), harnessing various types of intrinsic parallelism, are in active development. The main figures of merit to optimize are (F1) the computing power–the number of operations per second, (F2) energy efficiency–the energy consumption per operation, (F3) computing density–the number of operations per chip area per second, (F4) model scalability–the number of parameters supported by the optical system, and (F5) nonlinearity with compact footprint, low energy and latency.

Complementary-metal–oxide–semiconductor (CMOS) circuits, such as graphic processing units (GPUs) [13], tensor processing units (TPUs) [14], and application-specific integrated circuits (ASICs) [15], are the current workhorse in HPC. The high performance of CMOS processors derives from small feature sizes and a tailored system for memory communication, for instance, each input data is fed to $N$ processing elements (PEs) via electronic wires to improve the data bandwidth (Fig. 1b); however, the energy efficiency and clock rate scale inversely with the number of PEs due to wire capacitance ($E_w \propto CV^2$), which has become the fundamental bottleneck of CMOS electronics [16]. Quantum parallelism [17] is another attractive platform with exponential speedups in computation speed, but these systems are very tied to certain problems and their realization and scalability have been a long-term obstacle.

Optics is emerging with the promise to improve the computing figures of merit by many orders of magnitude, owing to nearly loss-less propagation and large optical bandwidth. Recent progress has shown matrix-vector multiplications (MVMs) using Mach-Zehnder-Interferometer meshes [18–20], phase change material (PCM) crossbars [21,22], or spatial light modulators [23–26] with the potential of speed-of-light computation [24,27,28], free-space data processing [25,26,29,30] and on-chip integration [18–21,23,31,34-35]. However, from the architecture level, existing techniques based on wavelength- or space-multiplexing, requires (1) $O(N^2)$ modulator devices due to mapping one modulator per matrix element; (2) high-speed analog-to-digital converters (ADCs) for readout that are energy-consuming; (3) high optical power for accumulating photon charges within each short readout times. Recently, time-multiplexed approaches have shown potential for large-scale computing but existing demonstrations based on optical interference [29,32] suffer from amplitude-phase coupling and interferometric fluctuations that limit the computing accuracy and scalability. From the device level, interfacing digital memory with high-speed, low-voltage, low-loss optical transmitters is necessary but still outstanding as a critical bottleneck, since existing technologies, such as microring-based silicon modulators require complex wavelength tuning and limited optical bandwidth, and broadband modulators suffer from trade-off limitations (in bandwidth, loss, and $V_\pi$), cannot fulfilled these requirements [33]. Innovations from architecture to device platforms are required for realize the optical advantages for efficient and scalable computing.

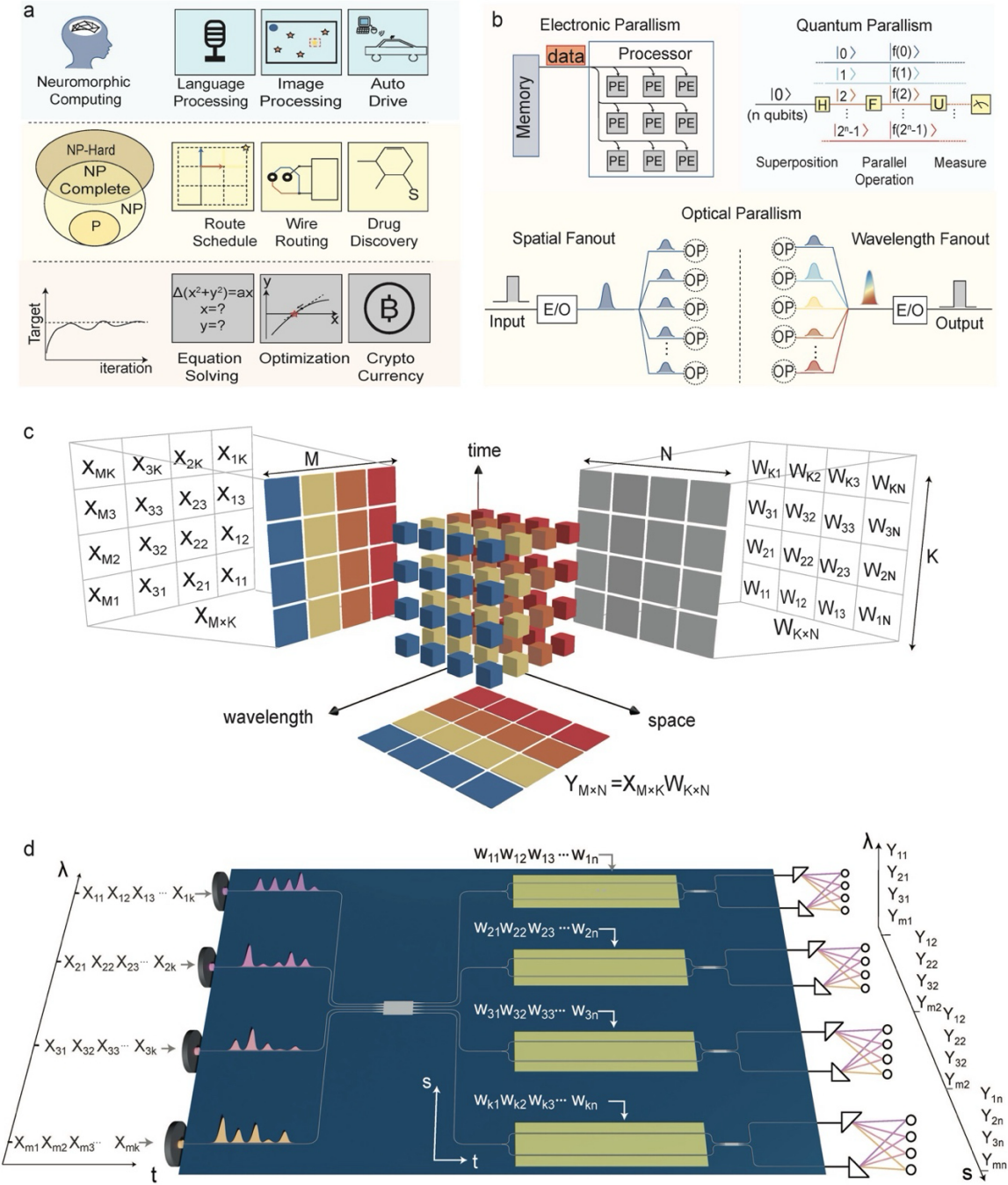

Fig 1. HITOP concept. (a) Tensor processing in HPC for neuromorphic computing, NP-hard solvers, and iterative solvers. (b) Comparison of various emerging computing platforms, utilizing data parallelism, quantum parallelism, and optical parallelism. PE: processing unit. H: Hadamard gate. F: operating function. U: unitary operator. OP: operation. (c) HITOP data format based on space-, time- and wavelength three-dimensional data encoding. The input matrix $X_{MxK}$ is encoded in time-wavelength dimensions with data fanout in space; the weight matrix is time-space encoded with fanout in wavelength. Dot products of overlapping elements are integrated over the time axis to lay out the MMM results in wavelength-space dimensions. (d). Chip-based architecture of HITOP showing 4 micron-scale laser wavelength channels for input matrix $X_{MxK}$ encoding and 4 spatial TFLN modulators for weighting.

Here we propose and demonstrate a hypermultiplexed integrated-photonics-based tensor optical processor (HITOP). With temporal data mapping, HITOP establishes neural connectivity with spatial beam routing and wavelength division multiplexing (WDM) (Fig. 1c), facilitating high density on-chip integration (Fig. 1d). It differentiates from state-of-the-art optical computing systems with its advantages of (1) scalability by using O(N) modulator devices for $O(N^2)$ throughput; (2) energy-efficient high-speed photonic material platforms with compact scalable laser modulators and broadband TFLN photonics for EO conversion (at 10s fJ/encoding); (3) high model scalability based on temporal data mapping (with each device activating 10 billion parameters per second); (4) time-integrating receivers to allow low optical power and scalable readout electronics; and (5) inline optoelectronic analog nonlinearity using laser threshold for latency reduction. All these advantages are crucial for scalable, efficient AI computing, as benchmarked with models of near half-million parameters and a potential full-system energy efficiency of 40 TOPS/W, representing >10-fold improvement compared to state-of-the-art digital systems.

## 2. Results

### HITOP architecture

HITOP is a hybrid optoelectronic system that performs general-purpose tensor processing $Y_{(M\times N)} = X_{(M\times K)} W_{(K\times N)}$ (Fig. 1c and d), where $M$, $K$, $N$ denote the matrix sizes (with $m$, $k$, $n$ being the matrix indices). Here the input matrix $X_{(M\times K)}$ is mapped to the wavelength-time bases, where the $m$-th vector (size 1x$K$) is amplitude encoded at the $m$-th laser wavelength with $K$ time steps, $E(\lambda_m, t_k) = X_{m,k}$. This wavelength-time encoding format has been exploited in our previous work in edge computing [36] for network model transmission. Here for on-chip tensor processing, we combine a total of $M$ wavelength channels via broadband multimode interferometers (MMIs) or wavelength division multiplexers. The combined beam is fanout with on-chip beam routing to $N$ spatial copies. At the $S_n$-th channel, the beam passes through a broadband optical modulator encoding a signed weight vector $W_n$ in $K$ time steps $E(t_k, S_n) = W_{k,n}$ to all the $M$ wavelengths. The cascaded modulation leads to multiplication on laser intensity $E(\lambda_m, t_k, S_n) = X_{m,k} W_{k,n}$. At the receiver, each $S_n$-th beam is wavelength demultiplexed on $M$ differential photodetectors, each subtract the output intensities of the two arms for operating positive and negative weight values (Methods). The resulting photocurrents are accumulated over $K$ time steps using a charge integrating amplifier, yielding

$$Y_{m,n} = \Sigma_{k=1}^{K} E(\lambda_m, t_k, S_n) = \Sigma_{k=1}^{K} X_{m,k} W_{k,n} \tag{1}$$

at the $m$-th wavelength of the $n$-th spatial output.

High scalability in HITOP is enabled by the system simplicity and low energy consumption. Matrix-matrix multiplication, which typically requires $O(N^3)$ MAC operations, is computed using O(M) wavelengths, O(N) modulators and O(K) time steps, where the number of optical modulators scales linearly with N+M. This simplicity is allowed by the high parallelism, where each input laser wavelength is spatially fanned out to all the modulators for simultaneous weighting, and each weight encoding applies to all the wavelengths using the broadband modulators. Similar to other device-based computing schemes, HITOP computes $O(N^2)$ multiply-accumulates (MACs) every time step, but it only reads out after integrating the MAC products over $K$ (≈1000) time steps, which significantly reduces the required optical power (Methods) to achieve the target computing precision, as well as the electronics power required for high-speed readout electronics.

*HITOP device platform*

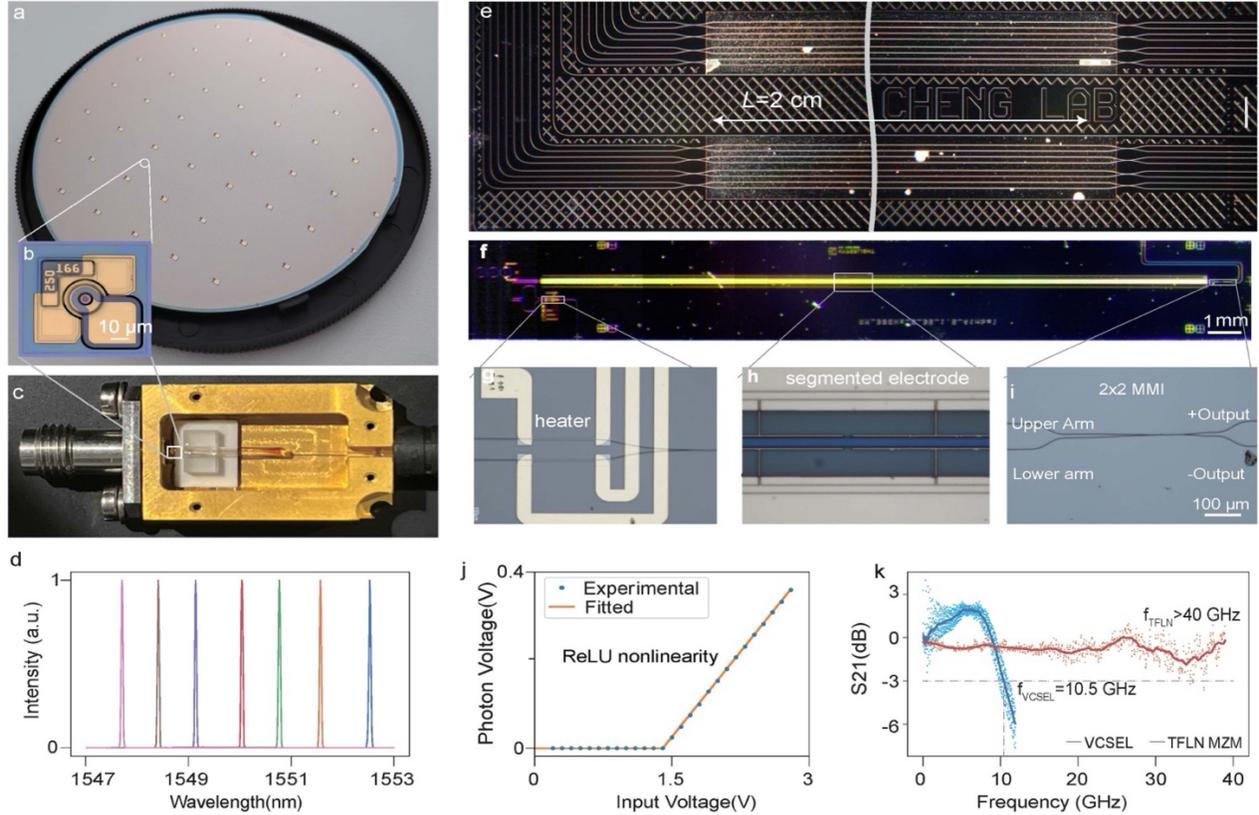

Fig 2. HITOP device platform. a. Wafer-scale fabricated VCSELs. A 3-inch wafer (VI system, Germany) with >100,000 single-mode VCSEL emitters, each with a 3 dB bandwidth >10.5 GHz . b: a single VCSEL laser with a footprint of 250x250 μm. c. A packaged VCSEL with electric wire bonds and fiber coupling. d. a spectrum with 7 VCSEL wavelengths used in HITOP, spanning 5 nm (0.7 THz). e. Dark view of a fabricated TFLN chip with 8 dual-port EO intensity modulators (one over eight is not functional), each modulator has 2 input ports and 2 output ports. The beam is coupled to one of the two input ports and the two output beams are subtracted to encode signed weights. f. An example of a 2-cm TFLN dual-output electro-optic intensity modulator. g. magnification view of heaters for setting operation bias. h. segmented electrodes for matching the microwave index and optical index [41]. i. dual-port outputs. j. VCSEL threshold effect for ReLU nonlinear activation between neural network layers. Data is measured with a photodetector with a TIA. A linear fit to the experimental data results in a statistical error of 0.26%. k. radio-frequency response of TFLN modulators with 3-dB bandwidth >40GHz and VCSELs with a bandwidth of 10.5 GHz.

We explore III/V-semiconductor VCSEL transmitters (Fig. 2b) for WDM, due to their scalability, high transmission speeds (>45 GHz [37]) and power efficiency (wall-plug efficiency (WPE) >57% [38]). Compared to WDM microring arrays for wavelength filtering and encoding [31], VCSELs are used for both laser generation and data transmission, where the cavity resonance and laser wavelength are naturally aligned without the need for complex biasing control and stabilization that are sensitive to environmental noises. Our VCSELs are fabricated at VI Systems GmbH (Berlin, Germany) with InAlGaAs/InP multilayer quantum wells as the active gain medium with emission wavelengths around 1550 nm [39,40]. A 300-mm wafer consisting of 100,000 VCSELs with a pitch of 250 μm is shown in Fig. 2a. Each VCSEL is packaged with an electronic wire bonded to an external driver for data modulation, and the output is coupled to a single mode fiber for WDM and edge coupling to the TFLN chips. The VCSELs are designed to emit at distinct center wavelengths around 1545 nm, 1550 nm and 1555 nm at room temperature, with a thermal tuning range of +/-2 nm to cover the whole spectrum of >10 nm. Seven VCSELs are tuned to match a standard telecommunication WDM grid (Fig. 2d), with a total optical bandwidth of 6 nm (0.7 THz) from (1547 nm to 1553 nm). The lasing threshold voltage is 1.2 V and each VCSEL emits >1 mW of optical power with a WPE of 20%. The 3 dB signal modulation bandwidth of each VCSEL exceeds 10 GHz (Fig. 2k). We verified the VCSEL data modulation accuracy by encoding a vector with normal distributed data. The response is recorded and compared to its ground truth, exhibiting an encoding error of <0.3% at 100 mega-symbol per second (MS/s), corresponding to 8~9 bits precision (Fig. S4). The accuracy decreases to 6 bits at 10 giga-symbol per second (GS/s) due to impedance mismatching, which can be addressed with custom driver circuit designs.

We developed a TFLN modulator platform (Fig. 2e) for weight streaming and time-wise dot products simultaneously with a broad range of wavelengths. TFLN photonics has emerged as a practical solution for optical transmitters to overcome current limitations in silicon photonics and combines the superior Pockels properties and scalable fabrication capabilities in next-generation integrated optoelectronic circuits [42]. A similar footprint, the performance of an EO-based Pockels modulator has overperformed the silicon or InP modulators in terms of lower switching voltage and optical loss (0.1 dB/cm achieved in our fabricated TFLN modulators), higher extinction ratio and EO bandwidth as well as better linearity, all of which are essential to HITOP devices with lower energy consumption, higher throughput and precision [33]. In addition, the thin film platform could be engineered with microwave electrode design to support ultrawide optical bandwidth (>500 nm, simulated and experimentally evidenced in supplementary Fig. S11 to Fig. S13), thanks to the tightly confined waveguides, precision nanolithography and compact device sizes as compared to bulk EO modulators. Our TFLN modulators are with traveling waveguide design, where microwave and optical index matching provides a >40 GHz electro-optical bandwidth with a $V_\pi$=1.3 V (2-cm length). The modulators are slightly unbalanced with a free spectral range >150 nm for the ease of characterization (Supplementary Fig. S3).

A key requirement to enable parallel processing is the design of dual-port TFLN modulators paired with differential detection to operate positive and negative weight values (Methods) that are crucial for AI computing. As encoding signed values is not straightforward with optical intensity, existing solutions [24,26,43] compute the positive and negative values separately in two time steps, which leads to computation overheads (in speeds, complexity and latency) and doesn't not allow simultaneously processing multiple vectors (Methods). Here in HITOP, each TFLN modulator is biased at the quadrature with equal power on two output arms that are canceled on differential detection (corresponding to zero values). Data modulation draws optical power from one arm to the other, generating positive/negative photon currents that are proportional to the encoding voltages. We sent in weight data with a peak-to-peak voltage of 0.6 V, corresponding to ~90 fJ/Symbol (Methods), and the low voltage operation keeps the modulators in a quasi-linear region, where we achieved <0.4% data encoding errors, corresponding to 8 bits of precision (Supplementary Fig. S4).

## HITOP benchmarking

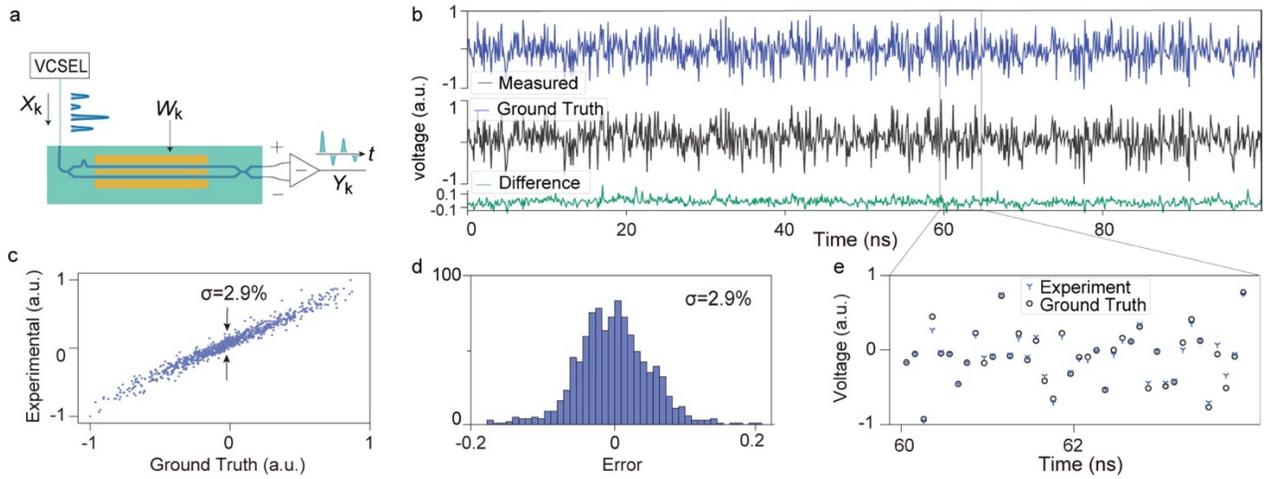

Fig 3. Experimental verification of HITOP computing accuracy at 10 GS/s. a. A vector computing unit consisting of a VCSEL and a modulator for cascaded modulation. b. Comparison of experimental results and digital ground truth of multiplying 2 random-distributed vectors with negative and positive weights. c. correlation of expected values and experimental multiplication results. d. histogram of the multiplication errors between ground truth and experimental results.

We constructed a computing system with 7 WDM VCSELs and 7 broadband TFLN modulators on a fabricated TFLN chip (Fig. 4a) (Methods). The HITOP computing accuracy is validated by sending two sets of data each with 1000 normally distributed random values to the VCSELs and the TFLN modulators (Fig. 3a). Noting that to achieve 6 bits operations at 10 GS/s, the delay between the *X* and the *W* synchronized pulse trains should be controlled within $\tau=100$ ps/$2^6$ = 1.56 ps, which is experimentally achieved by matching the electronic driver delay lines and optical delay lines before the differential detectors. Due to the high linearity of our transmitters, amplitude multiplication with signed weights via cascaded modulation achieves good consistency without any calibration required. The errors between the measurement and the ground truth exhibit a standard deviation of 2.9% at 10 GS/s (Fig. 3b-d) and 1.5% at 100 MS/s (Supplementary Fig. S8), corresponding to a computing accuracy of 5~6 bits. We attribute the 1-bit accuracy decrease at 10 GS/s to the electronic distortion and the timing mismatch in our measurement circuits. The achieved accuracy is sufficient for most AI applications.

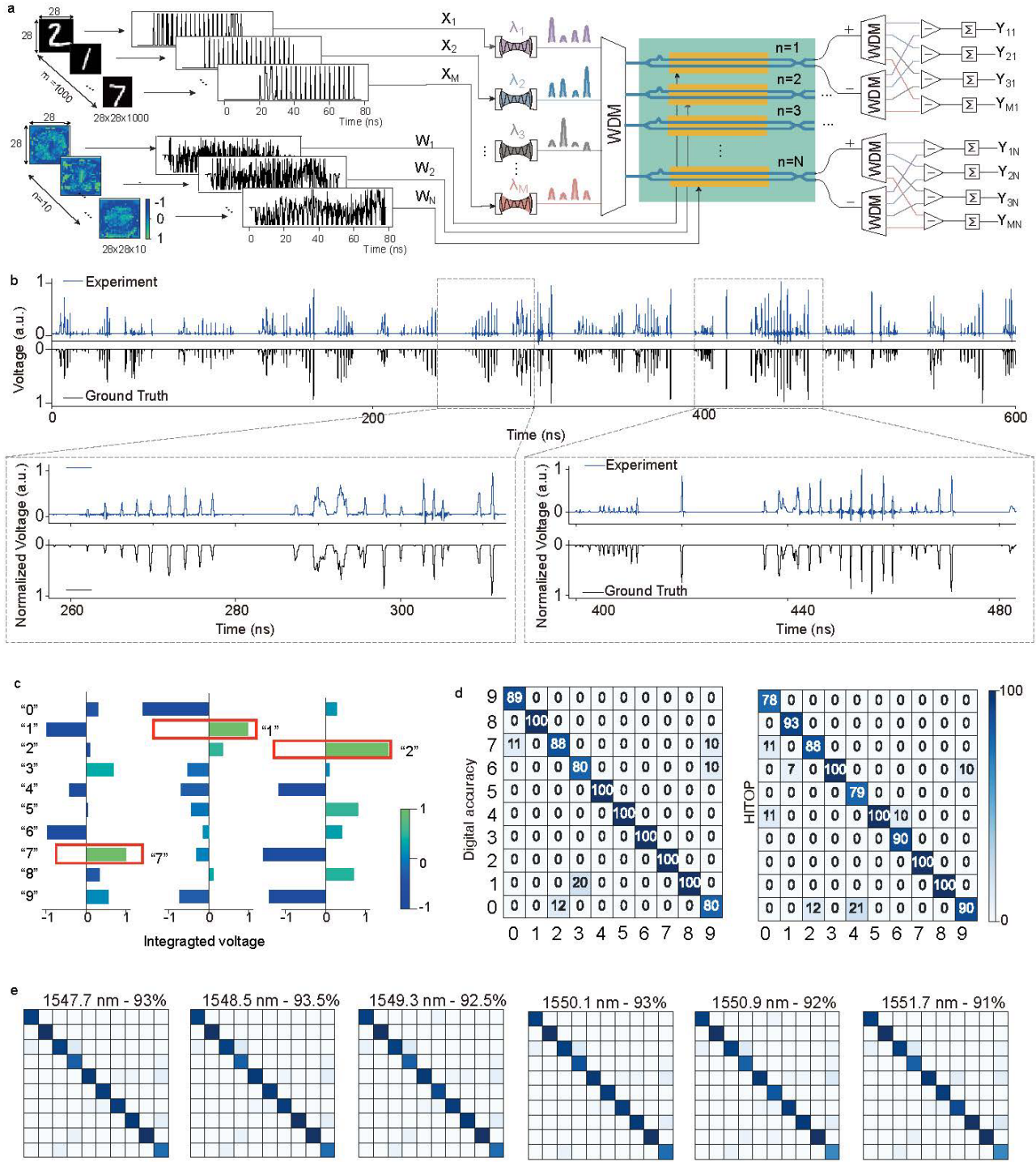

Fig 4. HITOP benchmarking with MNIST hand-written digit classification. a. Implementation of a single-layer pre-trained neural network model. The 10 weight vectors are flattened and streamed on the TFLN modulators. The images are encoded to VCSEL transmitters. b. a time trace multiplication result for digit classification with 10 GS/s data rate, the zoom-in figures show the time trace with great consistency to the digital multiplication. c. time-integrated voltages of 3 wavelength channels at the output of 10 spatial modes. d and e. Confusion matrix for MNIST digit classification at seven different wavelength channels showing good classification accuracy (with numbers in supplementary) as the digital model.

The HITOP computing accuracy, programmability and scalability are illustrated in AI inference tasks with image classification using the Modified National Institute of Standards and Technology (MNIST) datasets for handwritten digit, letter, and fashion recognitions. The computing accuracy is verified with pretrained AI models that require 4-5 bits precision (Supplementary Fig. S9) to achieve the accuracy. In experimental implementation, each image with 28x28 pixels, is flattened and converted into voltages modulating the VCSEL output intensity. With a one-layer network (28x28→10), where 10 weight vectors are encoded using TFLN modulators. We implemented an integrating charge amplifier (Methods) at each photo-receiver to accumulate the currents generated from each vector multiplication. The classification result for the digits is determined by which spatial channel that outputs the highest integrated voltage. With 10 GS/s data rate (Fig. 4), each image is processed within 78.4 ns. Without observing crosstalk, we achieved high classification accuracy at the seven wavelength channels with ~97% of the digital accuracy.

We reprogram HITOP for more complicated tasks with parallel processing and incorporate lasing ReLU inline nonlinearity. The VCSELs are biased below the threshold to activate the ReLU nonlinear function (hidden layers), which potentially allows cascading multiple analog layers without converting the partial sums to digital memories (Fig. S1). We verified the lasing threshold ReLU nonlinearity with VCSELs biasing at below threshold for random data encoding, which achieves (in Fig. S5-S7) accuracy of 2.78% at 10 GS/s, without observing bandwidth reduction due to crossing the VCSEL threshold. With limited high-speed electronic drivers available, we operated 4 channels simultaneously at 100 MS/s, and switching channels allowed us to collect data from all the device channels. We developed a three-layer network model (size 28x28→100→10) for recognizing Fashion images with ReLU function between the hidden layers. HITOP classifies over 91.8% among the 1000 random fashion image samples (Fig. 5), which is within the statistical error of the digital accuracy (94.3%). Moreover, the HITOP scalability is further benchmarked with the EMNIST letter dataset of 26 classification levels in a model size of 28x28→500→26. HITOP achieved 93.1% classification accuracy over 5000 images with a total of 4,050,000,000 operations, exhibiting excellent system stabilities. Noting that the model consists of 405,000 parameters, which is 25,000 more than previous integrated optical systems [18–22,34] (Table S4).

## *System performance*

HITOP suggests a path towards scalable optical computing with hyperdimensional parallelism. Its key advantages are illustrated with the computing figures of merit as well as its potential for future scaling.

(F1) The throughput (*T*) scales as T=2·*M*·*N*·*R*=0.98 TOPS, with M=7 VCSEL wavelength channels with N=7 TFLN modulators and R=10 GS/s. Though simultaneous operation of all the channels is limited by drivers available, this capability is verified with a combination of multiple channel operation at 100 MS/s and individual channels at 10 GS/s. Noting that the throughput scaling with O(*M*·*N*) at each clockcycle is achieved with O(N+M) modulators, which significantly simplifies system towards future scaling.

(F2) The energy cost of optical accelerators includes both the optical power and the electronic-optical-electronic (E-O-E) conversion. HITOP achieved high optical power efficiency at 18 fJ/OP experimentally (Methods), due to the advantage of integrating receivers. The fundamental requirement of optical power is set by the target computing precision, where less than <1 aJ/OP might be sufficient for 7-8 bits precision in standard thermal noise region (Supplementary Fig. S2) and this low optical power has been experimental benchmarking of <1 photon per operation [36]. This is crucial for scaling, as high on-chip optical power due to the waveguide damage and laser generation has become an ultimate limitation for optical systems.

The required electronic peripheral for driving HITOP is listed in Table S2 with energy cost from existing technology. For input E-O conversion, VCSELs and TFLN modulators encode data at 7 fJ/Symbol and 90 fJ/symbol, respectively (methods), which are further amortized by parallel processing with *N*=7 spatial channels and *M*=7 wavelength channels to the few femtojoule region. For output O-E conversion, integrating readout reduces the complexity of using high-speed analog-to-digital convertors (ADCs) that are energy consuming, e.g. CMOS photonic sensors [45,46] with charge amplification consumes <1 pJ/readout at at 10s MS/s. Besides, each readout consists of 2K (K=28x28) MAC operations, which converts the effective readout energy to fJ/OP. Therefore, with appropriate driving circuits, the full-system energy cost can reach ~25 fJ/OP (=40 TOPS/W) (Table S2), which is respectively, 40x and 500x better than state-of-the-art digital electronic computers (e.g, NVIDIA H100, Google TPU v4, GraphCore IPU2) and other integrated photonic systems [5, 7] (Methods).

(F3) The chip area density, $\sigma=2\times M\times R/A_2$=17.5 GOPS/mm$^2$, is currently limited by the size of the TFLN broadband modulators (20 x 0.4 mm$^2$) (Methods), as VCSEL modulators are compact for the same modulation speed and photodetectors are matured specially for low speed integrating receivers, e.g., 10s-million pixels in a imaging sensor. The computing density of TFLN modulators can be improved with more wavelengths for improved papalism and compact device designs.

(F5) Nonlinearity and latency. The ReLU nonlinearity with VCSEL threshold is based on analog E-O-E conversion. Without digitization, the latency is dominated by the time integration in charge amplifiers at K/R≈100 ns (at R=10 GS/s and K=1000), which is negligible for AI training and sufficient for most inference tasks that require real-time decision making.

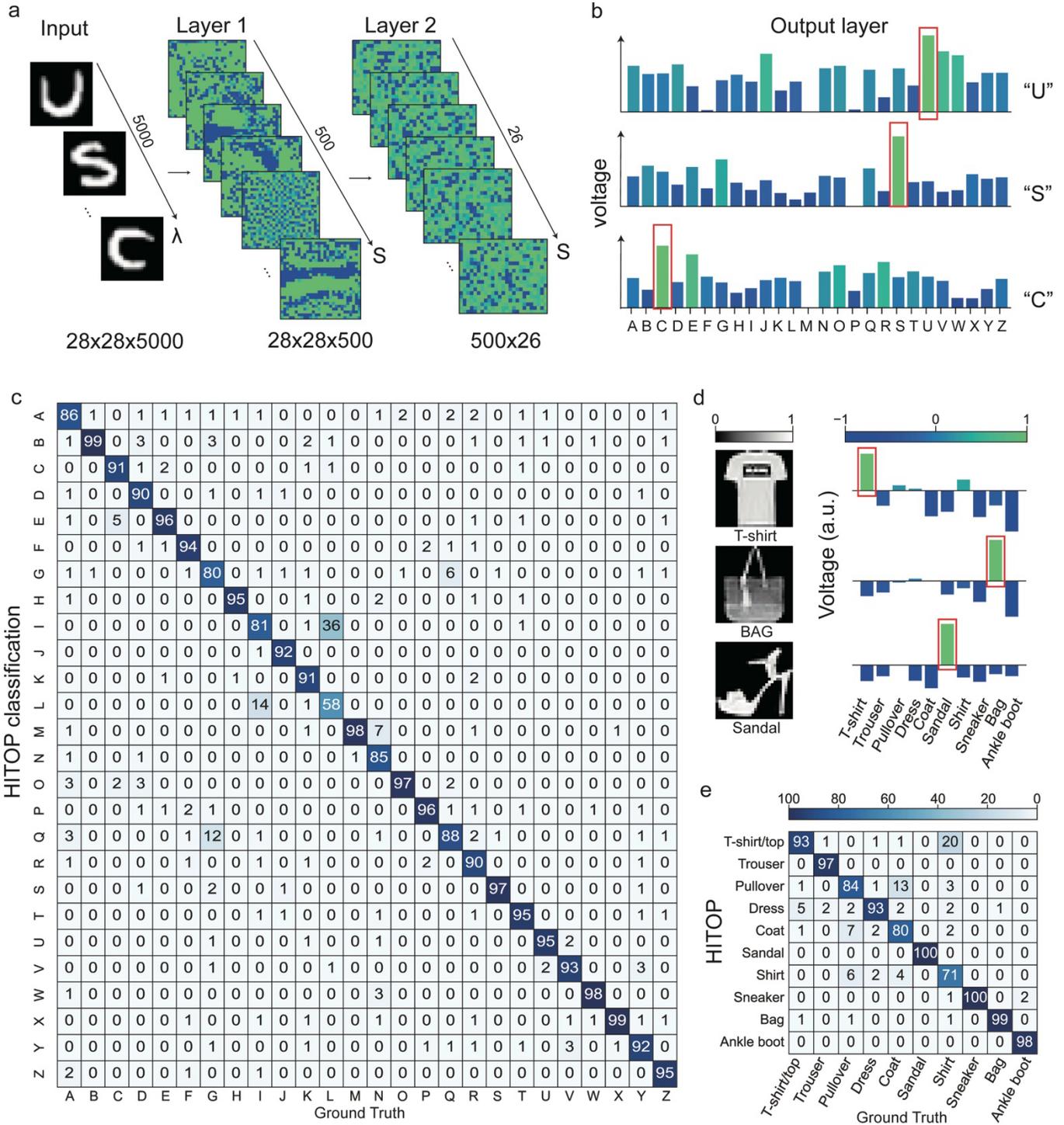

Fig 5 Experimental benchmarking of HITOP for different image classification tasks. a. Two hidden layers for the EMNIST letter classification with a network size of 28x28→500→26. b. The read-out voltages resulting from the HITOP output layer with high voltages for the letters "U", "S" and "C". c. Confusion matrix for EMNIST letter classification with a test set of 5000 images showing a classification accuracy of 93.1%. d. Fashion MNIST classification with a network size of 28x28→100→26. Experimental result of 3 example images. e. a confusion matrix for 1000 Fashion MNIST test images achieved an experimental accuracy of 91.8%.

## 3. Discussion

We explored the HITOP scalability, programmability, and energy efficiency for chip-based in-plane AI processing. Data mapping with temporal pulses reduces the number of encoder devices and enables future scaling in full system performance, e.g. with a total of 2000 modulators (i.e. M=1000 VCSELs and N=1000 TFLN modulators), HITOP can support $M \times N$=1000x1000 parallel channels with a throughput $T \propto 2 \times N \times M \times R$=20 POPS, which would otherwise require 1,000,000 modulators with device mapping schemes (e.g., MZIs [18], optical attenuators [34] or microring weight banks [31]). This high channel count may be achievable with existing technology as InP-based VCSELs covering wide frequency range over 1.3 µm-2 µm [48] and TFLN devices (with transparency window 350 nm-5200 nm [33,49–52]) with bandwidth over 500 nm ( Fig. S11-S13). Due to the low optical power required and the potential VCSEL-chip bonding in packaging (Methods), the VCSEL power (~6 mW per device) is sufficient for 1000 parallel computing channels at 6~7 bits precision (Fig. S2). The throughput of T=20 POPS is about 100x better than the best-in-class electronic systems (i.e., NVIDIA A100 or Google TPU v4 [14]), and moreover, the full system energy efficiency might reach sub-fJ/OP (1000 TOPS/W) with the high channels counts (Table S2), which is 1000x better than digital electronics. Further improvement with co-packaging optical digital-to-analog converters [53] may improve the efficiency further to few attojoules per operation.

HITOP with low energy cost and scalable computing will usher in a new era of possibilities and transformative applications across various domains. Based on three-dimensional data streaming with dynamic programmability, HITOP is suited for training models with trillions of parameters at low energy cost. This might enable larger AI models in language processing and machine vision and provide an eco-friendly solution with low latency to improve the processing power of smart, light-weight edge tensor processors for real-time decision making in autonomous vehicles and robots, and smart sensors in the internet of things. Beyond AI applications, the computing power will speed up simulation and modeling of complex physical and biological systems (e.g., climate changes and molecular dynamics) for accurate prediction of environmental challenges as well as drug discovery and healthcare.

## Methods

**TFLN Fabrication**

The TFLN MZMs are fabricated at Hyperlight Corp, the City University of Hong Kong and University of Southern California. The electrodes are designed as a traveling waveguide for RF signals with segmented fins for optics and microwave index matching [41]. Each MZM has a length of 2 cm and width of ~400 um with dual-port output. The MZM and MMI structure is patterned through electron beam lithography. Deep reactive ion etch (DRIE) is applied to shape the TFLN waveguide with an etch depth of 350 nm. A silicon dioxide is grown as the cladding layer. Annealing reduces the TFLN waveguide loss. A laser direct write lithography is applied to define the electrode pattern. Gold electrodes are deposited using electron beam deposition.

**Positive and negative weight encoding**

To encode signed values with optical intensities, solutions based on selecting and computing positive and negative values separately do not allow parallel processing because for element-wise multiplication, each weight vector requires sorting the corresponding input vectors differently and that can't be fulfilled simultaneously. So only 1 weight vector can be processed at a time. This is overcome with dual-port MZM designs for signed value encoding. The MZMs are designed to have two output ports for differential detection. Each MZM is biased at the quadrature point such that the input power is equally split to the 2 ports. Electro-optic data modulation draws optical power from one port to the other, creating a power difference that is proportional to the modulation voltages.

**Experimental setup.** The VCSELs are wire bonded and each beam is coupled to a telecommunication fiber (Fig. 2c). The emission wavelengths of 7 VCSELs are tuned to match a standard telecommunication WDM grid with a channel width of 100 GHz (OPTICO 100G DWDM 8CH MUX). The combined beams are split with fiber splitters and coupled to the TFLN chip with 7 modulators via edge coupling using standard C-band single mode fibers. The data in Fig. 4 were taken at 10 GS/s (Tektronix AWG70002B) with driving 1 VCSEL a 1 TFLN modulator with synchronization and delay matching. The multi-channel operation (Fig. 5) were demonstrated with two wavelengths and two TFLN modulators at 100 MS/s (Tektronix AWG5014C), moving the electronic probes and drivers to different channels allows us to incorporate more devices. At the output of each modulator, the wavelengths are demultiplexed and detected with differential detection (Thorlab PDB450C). The generated photon voltage is sampled by a high-speed oscilloscope (Rohde & Schwarz RTO6 for 100MS/s and RTP134B for 10GS/s). In terms of the model benchmarking, the photon voltage is integrated using time integrating receivers and digitized with an FPGA-based data acquisition board (Alazar Tech 9416).

**Time integrating receiver**

The time-integrating receivers are homemade with a charge-integrating amplifier (Texas Instruments IVC102) soldered at the backend of the photodetectors [36]. The integration is triggered on for a target integrating time steps and switched off when the trigger voltage is set to zero. When time integration is on, the capacitor accumulates photocurrents over time and converts the integrated charges to photo-voltage that can be digitized.

*Neural network training*

We trained our neural network models with the standard PyTorch package. The model consists of one input layer, two fully connected hidden layers, and an output layer (Fig. 4). The input layer consists of 784 neurons, corresponding to the 28x28 pixels of a full-size MNIST digit and letter fashion MINIST image. The input is passed to two fully-connected hidden layers, with matrix-matrix multiplication (batch operation) to speed up the process. For digit MNIST and Fashion MNIST, the output layer consists of 10 neurons, each neuron represents a digit (from 0 to 9) or a fashion style. For the letter MNIST, the output layer consists of 26 neurons, corresponding to the 26 levels (from a to z) of classification. The prediction result of which digit is given by the number of neurons which has the largest value. We utilize the cross entropy loss function and retrieve the gradients in each iteration in a model implemented in PyTorch. A random additional noise is introduced during the training process to enhance the robustness of the model. A large learning rate is set to start the training and is gradually reduced to optimize the accuracy. All the models are trained using Adam as the optimizer and the cross-entropy function for the loss evaluation. The model parameters are listed in Supplementary table S3.

**Neural network benchmarking.** We deployed a 3-layer neural network model with a model size of 28x28→100→10 for MNIST digit classification and fashion classification. ReLU nonlinearity based on the lasing threshold is incorporated between two layers. Running a test set consisting of 1000 random hand-writing digit images yielded a classification accuracy of 95.5%, which is close to the model accuracy of 95.8%. For Fashion MNIST, which is a more challenging task due to the higher image complexity, the experimental accuracy of 91.8%, which is 97.3% of digital accuracy (94.3%). With a larger model size (28x28→500→10) for EMNIST letter classification, an experimental accuracy of 93.1% is demonstrated compared to the digital accuracy of 93.4%.

**Potential system performance.**

*Energy cost of devices.* Our VCSELs are biased with a DC voltage $V_{VCSEL-DC}$=1.3 V with a current $i_{th}$≈2mA, which is at a near-threshold current region with 2.6 mW DC output optical power. A positive RF signal with peak-to-peak voltage of 0.6 V is sent to VCSEL with a resistance R=650 Ω consuming 70 µW for data encoding, which is equal to 7 fJ/Symbol at 10GS/s. The TFLN MZMs are biased at the quadrature point and driven within the quasi-linear region with peak-to-peak voltage $V_{pp}$=0.6 V ($V_\pi$=1.3 V), which yields a root-mean-square drive voltage $V_{rms}$=0.2 V. The AC signal is encoded with 50 Ω termination. Its energy per symbol is estimated as $P_{MZM}=V_{rms}^2/(R \cdot 50$ Ω )=88 fJ/Symbol. The energy cost of the potential electronic drivers is listed in Table S2.

*Compute Density.* Compared to the compact VCSEL devices ($A_1$=0.06 mm² per device), the broadband TFLN modulators ($A_2$=8 mm² per device) dominate the chip area, with a computing density of $\sigma_0$=2×M×R/$A_2$=17.5 GOPS/mm² with M=7. In future development, with M=1000 and compact device design ($A_2$=2 mm² per device), the area efficient is expected to reach $\sigma_1$=2×M×R/$A_2$=10 TOPS/mm².

<u>Optical power budget</u>: The time integrating accumulates photocharges over K time steps to provide a higher signal-to-noise ratio that sets the computing bit of precision. We calculate the power budget for a target SNR of 100 (6~7 bits) by including the detector noise, laser intensity noise, and shot noise (Supplementary Information 2). With K=1000 (28x28 in experiment), HITOP requires only 0.3 µW optical power on each PD (NEP=2 pW·Hz$^{-½}$), which sums up to 0.3 W for 1000x1000 channels on chip power. This can be provided with 1000 VCSELs (each with 0.3 mW). Taking into account the loss of VCSEL-to-chip coupling (5 dB demonstrated in [44]) and the TFLN modulator loss (0.1 dB/cm), about 1 mW per VCSEL is sufficient. With 1000 wavelength channels in future scaling, the total on chip power is ~ 1 W. This number can be reduced to 30 mW with K=10$^6$ (Fig. S2).

**Potential system packaging.** HITOP supports heterogeneous integration with optoelectronic packaging. Existing photonic integration technologies, such as VCSEL-waveguide bonding [44], WDMs based on microring filters [54], heterogeneous integration of photodetectors on TFLN [55,56], can be exploited to integrate the whole system on the same chip. An electronic integrated circuit (with pitch-matching DACs, charge integrating amplifiers, and ADCs) will be flip-chip bonded to the photonic chip. Since our photonic devices operate with voltages <1 V, the electronic circuit is compatible with standard CMOS technology, where a chip area equivalent to the VCSELs ($A_1$=0.06 mm$^2$) and a TFLN modulator (8 mm$^2$) can store, respectively, ~60,000 and 47-million 8-bit digital values for input buffer and weight storage, as a static random access memory (SRAM) cell sizes only 0.021 μm$^2$ in 5-nm bulk CMOS chip [62]). The same footprint of a TFLN modulator also supports more than one billion transistors for digital logic, DACs, TIAs, ADCs.

**Acknowledgments:** This work is supported by the DARPA NaPSAC program under project No. N66001-24-2-4002. Z.C. and M.Y. acknowledge the Optica challenge awards. Z.C. thanks Dr. Nikolay Ledentsov Jr. from VI Systems for supporting the VCSEL samples and useful discussion. D.E. thanks Army Research Office under grant number W911NF17-1-0527, and NTT Research under project number 6942193 and the NTT Netcast award 6945207.

**Author contributions:** Z.C. and M.Y. conceived and supervised the project. S.O., K.X. and L.Z. performed the experiment and finalized the experimental results, assisted by A.S. and R.H.. A.S. developed the integrating receivers. S.O. and E.Z. created the software model for neural network training. K.Z., H.F., and C.W. fabricated the TFLN MZMs. R.K. and C.L. performed simulation on TFLN modulators. D.E. provided the experimental support and discussions. Z.C. wrote the manuscript with contributions from all the authors.

**Competing interests:** Z.C, S.O. and M.Y. have filed a patent related to HITOP, bearing application 064189-0980(2023-049-01). Z.C., R.H, and M.Y are co-founder of Opticore Inc. and hold equity. D.E. serves as scientific advisor to and holds equity in Lightmatter Inc and Opticore Inc. A.S. is a senior photonic architect at Lightmatter Inc. and holds equity. Other authors declare no competing interests.

**Data and materials availability:** Data and materials are available from the corresponding author upon reasonable request.

# Supplementary Information

## *1. Scheme of HITOP Multi-layer implementation*

HITOP is an optoelectronic architecture. The input and weights are stored in nearby memories and transmitted to the optical system for matrix operations in the optical domain. The multiplied result is read out with charge accumulation and serialized with nonlinearity activation to the next layer. With a memory unit integrated together, our ONN could be used for multiple layer neural network inference with such working flow:

● Input vectors and weight vectors for the current layer are read out from the digital memory and programmed as time traces.
● Each input and weight time trace, respectively, is sent to the corresponding VCSEL or TFLN MZM (Fig. 1d).
● The output is acquired and accumulated by the photodetectors and time integrator arrays.
● Digital memory stores the spatially distributed signals from the photodetector and time integrator array.
● Reprogram the output by mapping the spatial distributed signal into corresponding time steps. The outputs of the same wavelength at different spatial channels are serialized as the input driving signal to a VCSEL transmitter in the next layer. Here the ReLu nonlinear operation function is encoded using the lasing threshold effect.

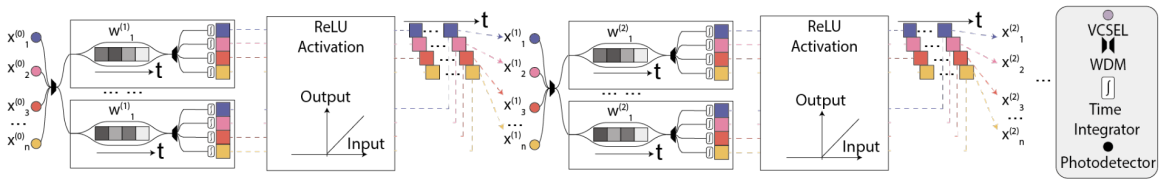

Fig. S1 HITOP forward propagating deep neural network architecture. The network consists of an arbitrary number of layers (2 layers are shown). Each layer computes a matrix-matrix multiplication in wavelength-time-space multiplexing using VCSELs and TFLN MZM (Fig. 1d).

## 2.  Optical power budget of time integrating receivers

The total noise is a combination of detector noise, photon shot noise and laser intensity noise. The signal to noise ratio analysis was studied in [65,66], validated in our previous work for optical computing [29] and adapted here. The total noise spectral density is

$$N_{tot} = \sqrt{[(\eta\alpha)^2 + 2h\nu\eta P + \beta(\eta P)^2]B}$$

where α is the noise equivalent power (NEP) of the detector, η is the detector quantum efficiency, h is the Planck's constant, ν is the laser frequency, P is the laser power, and β is the relative laser intensity noise. The noise spectrum is integrated over the effective Fourier bandwidth of B=1/(2T), where T is the acquisition time. With the root square mean input signal $S = 2\sqrt{2}\eta P$, where a factor of 2 enhancement of the signal account for the differential subtraction using 2 photodetectors in HITOP, the average uncertainty of the signal is determined as

$$\sigma_t = N_{tot}/S = 1/(2\sqrt{T})\sqrt{[(\alpha/P)^2 + 2h\nu/(\eta P) + \beta]}$$

So the total signal to noise ratio is

$$SNR = 1/\sigma_t = 2\sqrt{K/R}[(\alpha/P)^2 + 2h\nu/(\eta P) + \beta]^{-1/2}$$

where T=K/R, K is the integration time steps, and R is clock cycle.

To reach a SNR of 100 (for 6-7 bits of computing precision) at 10 GS/s, a optical power of P=40 µW is required at each PD, with the detector noise of NEP=2 pW·Hz$^{-½}$, and P=120 µW for NEP=10 pW·Hz$^{-½}$, RIN=-135 dBc/ limited by the laser intensity noise at this power level (Fig. S2a). Considering the high speed detectors are relatively noisy (due to limited chip area and quantum efficiency), the NEP=2 pW·Hz$^{-½}$ is practically hard to achieve. So the total power requirement when the system scales up to 1000x1000 channels is $P_{tot}$=120 W (and P=400 W with ~5 dB loss due to VCSEL-chip coupling [44]), which is challenging for on-chip laser generation and power handling of photonic circuits.

This power requirement is significantly reduced with time integration. Although the computation clockrate is high (e.g. 10 GS/s), the integrating receiver only digitizes the vector products when the accumulation is complete (e.g. 10 MHz with vector size K=1000, or 10 kHz with K=10$^6$). The slow readout detection allows (1) sufficient measurement time to accumulate photon charges, as we previously demonstrated less than 1 photon per MAC operation in Ref. [36]; (2) usage of low-speed photon detectors of high sensitivity. In the following plots, we calculate the SNR evolution with increasing optical power under our experimental conditions of the PD sensitivity (noise equivalent power NEP=2 pW·Hz$^{-1/2}$ including amplifier) and K=1000. Our system only requires 0.3 µW optical power on each PD, which sums up to 0.3 mW for 1000 channels. Taking into account the loss of VCSEL-to-chip coupling (5 dB demonstrated in [44]) and the TFLN modulator loss (0.1 dB/cm), a total power of about 1 mW per VCSEL is required. This number can be reduced to 0.03 mW with K=10$^6$. Therefore, our VCSEL power of 6~8 mW is sufficient to support the 1000 channels without amplification required in either case.

For the detector NEP=10 pW·Hz$^{-½}$. The power requirement for K=1000 increases to 1.6 µW per PD (1.6 mW for 1000 channels), which is considered the limit of our VCSELs for 1000 channels (Fig. S2d). Again, this requirement is lower by a factor of ~30 (SNR∝T$^{-1/2}$) to 50 nW per PD (50 µW for 1000 channels), when the integration time is K=10$^6$ (Fig. S2f).

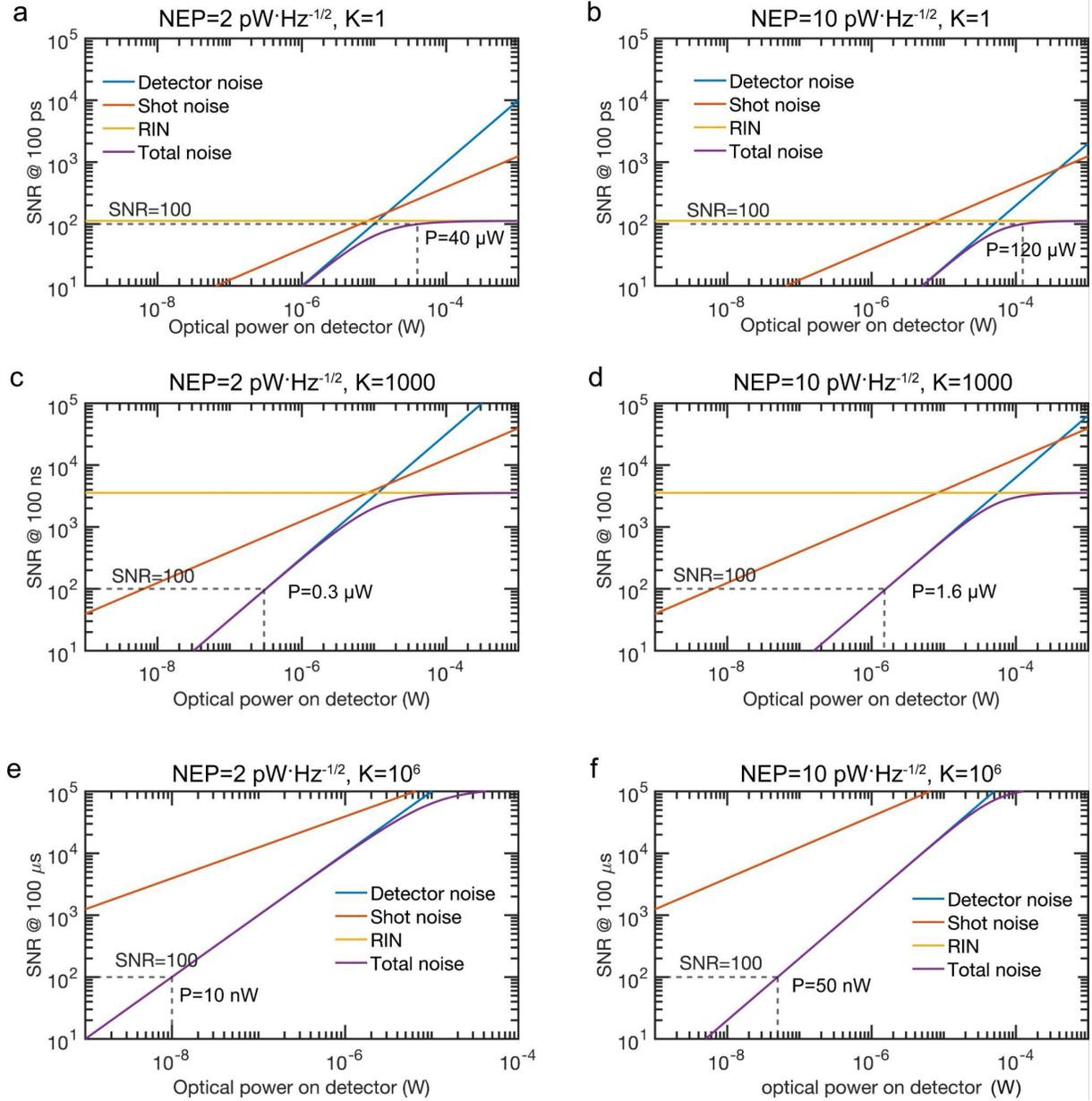

Fig. S2 the evolution of signal-to-noise ratio with input optical power. The power/PD requirement is labeled to reach a SNR=100 at each scenario. The SNR is calculated at the laser frequency of 195 THz, $\eta$=0.9, R=10 GS/s and laser RIN ($\beta$)=135 dbc·Hz$^{-1/2}$ a. the detector noise equivalent power NEP=2 pW·Hz$^{-1/2}$, acquisition time T=K/R=100 ps. b. NEP=10 pW·Hz$^{-1/2}$, T=K/R=100 ps. c. NEP=2 pW·Hz$^{-1/2}$, K=1000, T=K/R=100 ns. d. NEP=10 pW·Hz$^{-1/2}$, K=1000, T=K/R=100 ns. c. NEP=2 pW·Hz$^{-1/2}$, K=10$^6$, T=K/R=100 μs. d. NEP=10 pW·Hz$^{-1/2}$, K=10$^6$, T=K/R=100 μs.

Table S1 Power budget for different experimental conditions

| Power budget | Detector NEP=2 pW·Hz$^{-½}$ | | | Detector NEP=10 pW·Hz$^{-½}$ | | |
|---|---|---|---|---|---|---|
| | K=1 | K=1000 | K=10$^6$ | K=1 | K=1000 | K=10$^6$ |
| Power/detector | 40 µW | 0.3 µW | 10 nW | 120 µW | 1.6 µW | 50 nW |
| Power per laser for 1000 fanout (with 5 dB coupling loss [44]) | 120 mW | 1 mW | 30 µW | 400 mW | 50 mW | 150 µW |
| Total optical power for 1000x1000 channels | 120 W | 1 W | 30 mW | 400 W | 50 W | 150 mW |

## 3. Full system energy consumption and electronic peripheral

HITOP is an optoelectronic processor. The power consumption of drivers and electronic components is listed below:

Table S2 Energy consumption of electronic driving circuits of HITOP

| Components | Energy budget | Parallelism[b] (Now) | Energy/OP (Now) | Parallelism (near-term development) | Energy/OP (near-term development) |
|---|---|---|---|---|---|
| VCSEL DC driving | 2.6 mW | 2N=14, R=10 GS/s | 18 fJ | 2N=2000, R=10 GS/s | 0.13 fJ |
| VCSEL AC driving | 7 fJ/use | 2N=14 | 0.5 fJ | 2M=2000 | 3.5 aJ |
| TFLN modulator AC driving | 80 fJ/use[a] | 2M=14 | 5.7 fJ | 2M=2000 | 40 aJ |
| Integrating photoreceiver (detector+capacitor+TIA) | 1 pJ/use [45] | 2K=28x28x2 | 0.6 fJ | 2K=2x10$^6$ | 0.5 aJ |
| ADC at 10 MHz[c] | 1 pJ/use[63] | 2K=28x28x2 | 0.6 fJ | 2K=2x10$^6$ | 0.5 aJ |
| Total | | | 26 fJ/OP | | 45 aJ/OP |

[a]TFLN modulator energy is calculated from the peak-to-peak driving voltage of 0.6 V with the 50 Ω load (Methods).
[b]The factor of 2 in parallelism accounts for the multiplication and accumulation operations.
[c]The system only readout once after integrating over 1000 time steps.

## 4. TFLN Device characterization

Our MZMs are designed with sight imbalance path length between two arms, for the ease of calibration, which is not necessary in future full-system development. The free-spectral range is >150 nm, which is negligible in our demonstration with a total bandwidth of <5 nm.

The response of lithium niobate amplitude modulation is a sinusoidal function with increasing driving voltage. When encoding positive and negative weight values, the modulator is biased at the quadrature point with driving voltage of ($V_{pp}$=0.6 V), which is quasi-linear. Without calibration, the encoding accuracy is 6 bits precision due to the slight nonlinearity. The slight nonlinearity at close extinction point can be pre-calibrated using an inverted transfer function to reach a linearity up to >99.6%, which corresponds to 8 bit precision.

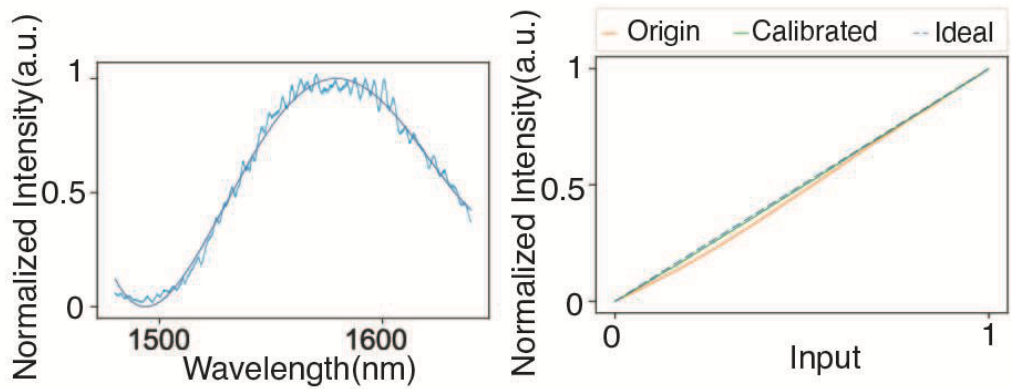

Fig. S3. TFLN MZM characteristics, left: the free spectrum range of a quasi-balance arm MZM, right: Linearity of the transfer function of the MZM, without calibration, we achieved 6 bits of encoding precision. With calibration, the standard deviation of the residue between the calibrated transfer function and the ideal transfer function is less than 0.4%.

# 5. Data Encoding Accuracy

We investigate the encoding accuracy of our opto-electro transceiver, i.e. VCSEL and TFLN modulator, by applying random data distributed in the range of [0,1]. Due to the linearity of VCSEL response, when driving in the near-threshold region, we achieved 8-bit precession without extra calibration at low clock rates. With the linear calibration of MZM response, we also achieve 8-bit precession encoding accuracy with a single output TFLN MZM. Limited by our photodetectors and the measurement electronics, the encoding accuracy drops to 6 bits when the clock rate goes up to 10 GHz.

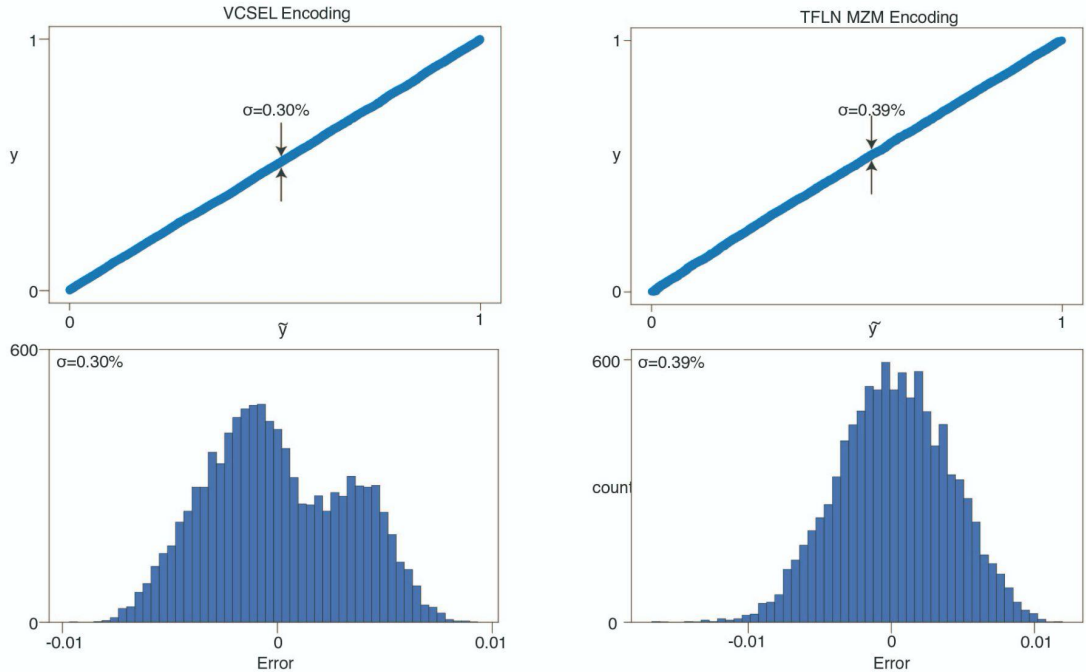

Fig. S4. Input-output voltage response accuracy of the VCSEL encoding and residuals (left) and TFLN MZM encoding (right) showing 8-bit precision.

We verify the VCSEL response accuracy when biasing around or slightly below the lasing threshold (Fig. S5 ~ S7). The VCSELs are biased at $V_{dc}$=1.2 V and the image data were flattened and encoded with a peak voltage of $V_{ac}$=0.6 V (same as our experimental conditions). The encoding results match well with the ground truth. For an example image "0", we achieved the encoding errors of 2.26%, 2.3% and 3.11%, respectively, at the data rate of 1 GHz, 5 GHz, and 10 GHz. The decrease in accuracy at 10 GS/s is due mainly to the impedance of drivers and VCSELs, which leads to electronic interference ripples.

The encoding of image "1" and "3" at 10 GS/s exhibit errors of 2.3% and 2.5% respectively (Fig. S6). An average error of 2.78% over the 10 consecutive images (Fig. S7) is observed, corresponding to 5~6 accuracy, which is sufficient for our computing tasks. We didn't experience degradation of accuracy due to the power and spectrum instabilities.

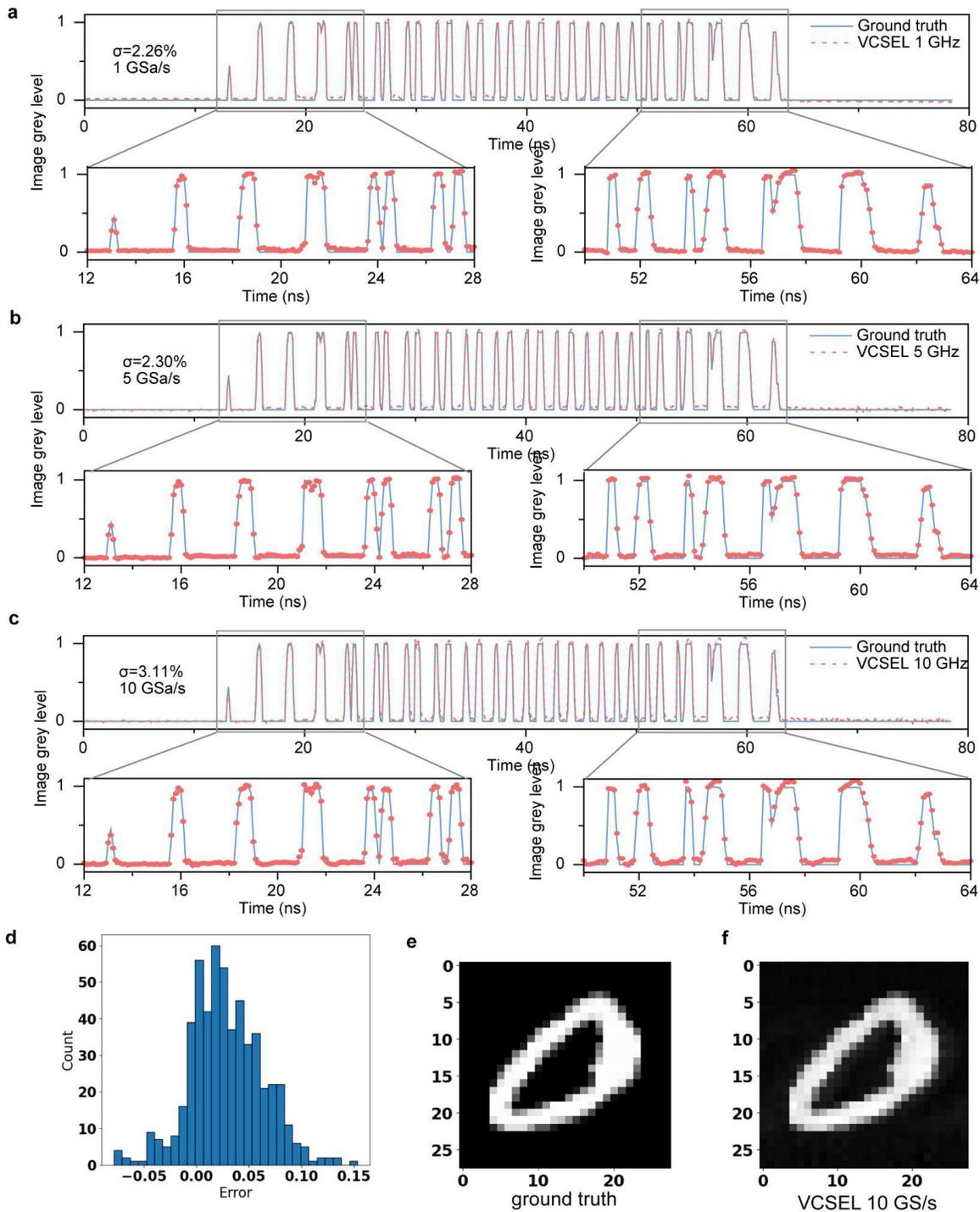

Fig. S5 Verification of image encoding accuracy with VCSELs biasing slightly below lasing threshold for ReLU nonlinearity. **a-c**. The encoding time trace and statistical errors achieved 2.26%, 2.30% and 3.11% at the data rates of 1 GS/s, 5 GS/s and 10 GS/s, respectively. d. error distribution at 10 GS/s. e-f. comparison of VCSEL encoded image with ground truth.

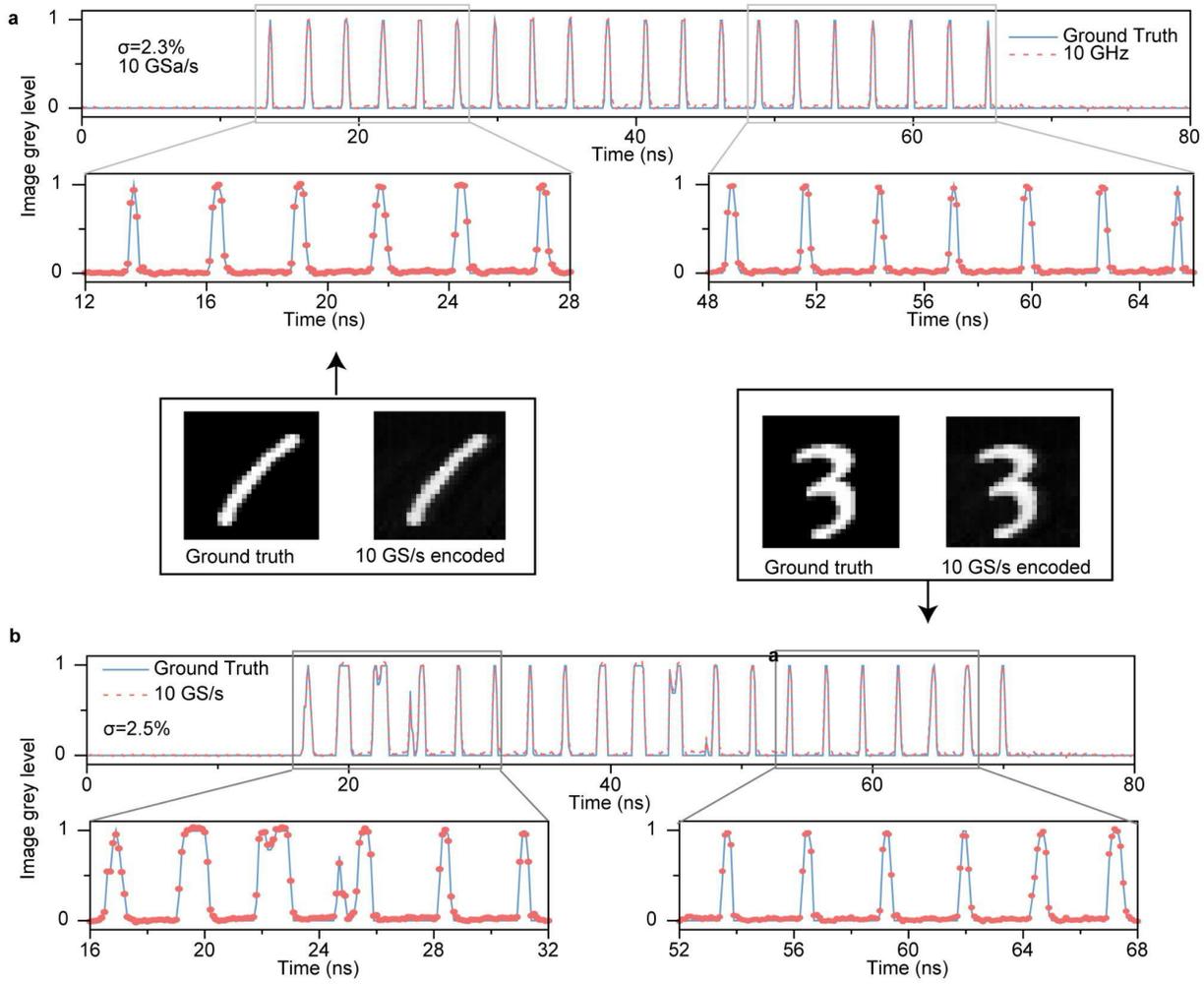

Fig. S6 examples of VCSEL encoding random digit images "1" and "3" using VCSEL at the data rate of 10 GS/s. The result shows 2.3% and and 2.5% errors compared to their ground truth.

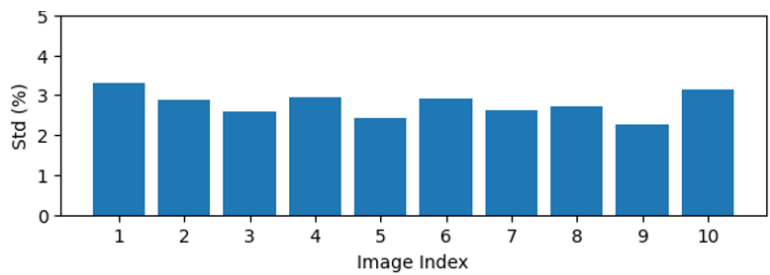

Fig. S7. VCSEL encoding accuracy for consecutively 10 images achieved an average statistical error of 2.78%.

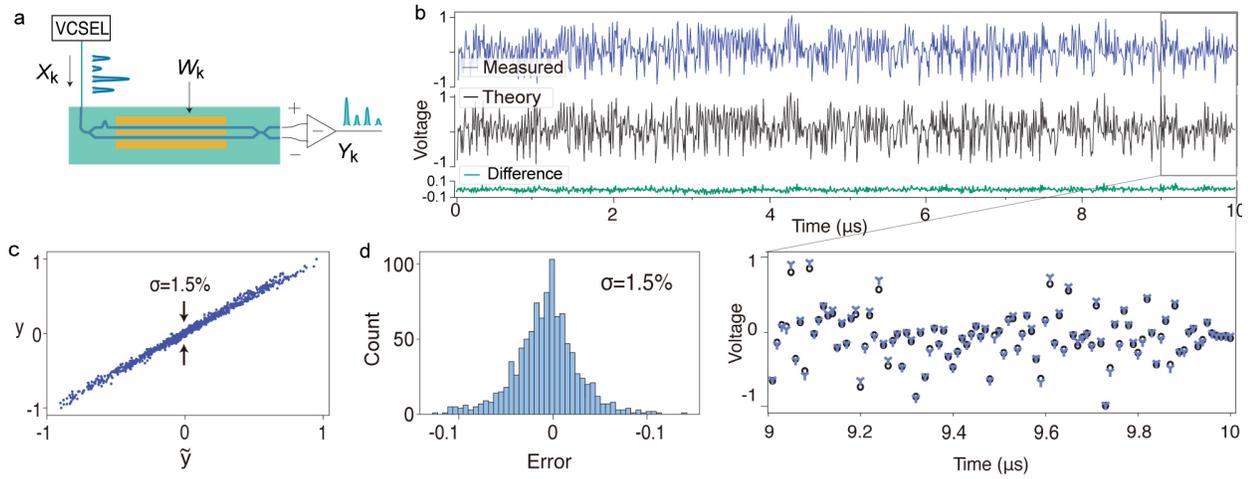

Fig. S8 Experimental verification of HITOP computing accuracy at 100 MS/s. a. A vector computing unit consisting of a VCSEL and a modulator for cascaded modulation. b. Experimental results of multiplication of 2 random-distributed vectors with negative and positive weights. c. correlation of expected values and experimental multiplication results. d. histogram of the multiplication errors between ground truth and experimental results.

## 6. Neural network models for different tasks

To benchmark the performance of our HITOP system, we trained several neural network models for 3 different datasets. All the models are built using PyTorch. Adam optimizer is deployed for training all these models with different electronic hardwares (CPU, Nvidia GeForce 3070, Nvidia Tesla V100). During the training process, all the parameters are restricted within a range of [-1,1] to match our HITOP system. Additionally noise is applied to enhance the robustness of the model. The hyperparameters and the corresponding digital and experimental accuracy are listed in Table S2.

In addition, we analyzed the required computing precision to implement our AI models. We observed the statistical accuracy of our image classification, by quantizing the output product of each layer to be B=1 bit, 2 bits, 3 bits, …, 16 bits, respectively, where the signals are normalized and before approximating to the closest level of $1/2^B$. The results are listed in Table S3 and plot in Fig. S4. The fact that our experimental results are close to the maximal digital accuracy indicates that our computing accuracy is higher than 4~5 bits.

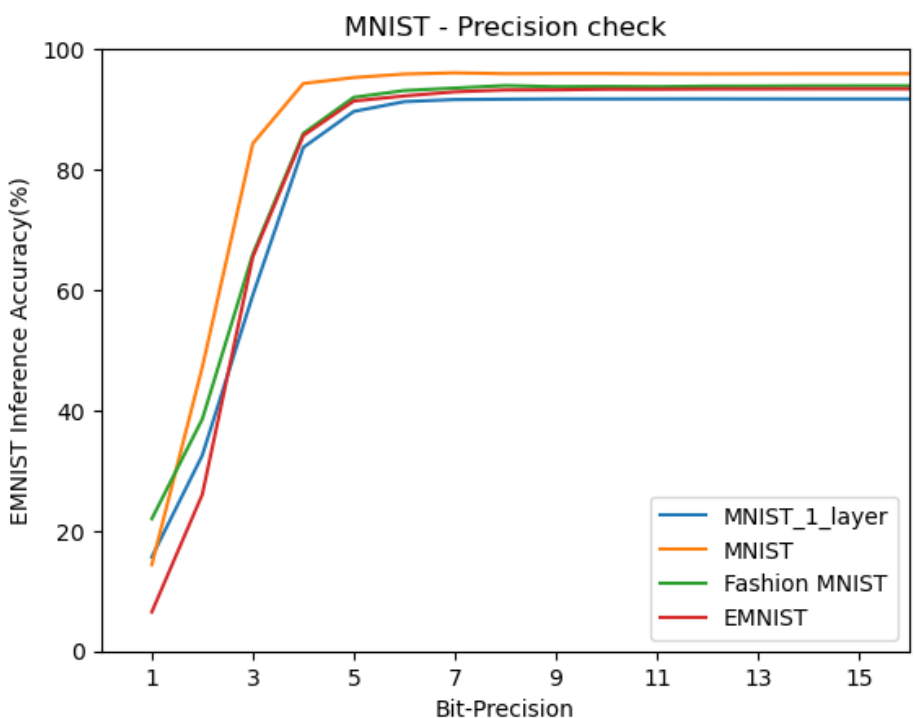

Fig. S9 Model accuracy at different quantization bit precision. The model accuracy increases with higher computing bit precision and reaches maximum at 4~5 bits of precision.

Table S3. The accuracy of our AI models at different computing precision.

| Bit-Precision | MNIST_1_layer | MNIST | FashionMNSIT | EMNIST |
|---|---|---|---|---|
| 1 | 15.6% | 14.4% | 22.0% | 6.5% |
| 2 | 32.5% | 47.2% | 38.6% | 26.0% |
| 3 | 59.1% | 84.3% | 65.9% | 65.5% |
| 4 | 83.7% | 94.3% | 86.0% | 85.6% |
| 5 | 89.6% | 95.2% | 92.0% | 91.4% |
| 6 | 91.2% | 95.9% | 93.1% | 92.2% |
| 7 | 91.6% | 96.0% | 93.5% | 92.9% |
| 8 | 91.7% | 95.9% | 93.8% | 93.2% |
| 9 | 91.7% | 95.9% | 93.8% | 93.3 |
| 10-16 | 91.7% | 95.9% | 93.8% | 93.4% |

Table S4. Model size and demonstration of integrated optical systems

| References | dataset | Input data size | Number of effective parameters in optics | Exp. accuracy |
|---|---|---|---|---|
| Shen 2017 [18] | Vowel | 1x4 | 4x4 =16 | 76.7% |
| Feldmann 2021 [21] | MNIST | 28x28 pixels | 4x2x2 =16 | 95.3% |
| Ashtiani 2022 [34] | 4-level EMNIST | 3x4 pixels | 3x4→3→2 =40 | 89.9% |
| Bandyopadhyay 2022 [19] | Vowel | 1x6 | 6x6x3 =108 | 92.7% |
| Pai 2023 [20] | moons | 64 pixels | 4x4x3 =48 | 97% |
| Zhu 2022 [35] | MNIST | **28x28 pixels** | **1x10** =10 | 91.4% |
| | Fashion MNIST | | | 80.4% |

| | | | | | |
|---|---|---|---|---|---|
| Mourgias-Alexandris 2022 [64] | MNIST | **1x4 (two hidden layers in a CNN)** | **1x4→1x2** | **=6** | **>97%** |
| Dong 2023 [22] | ECG dataset | 1x35 | 3x1x3 | =9 | 93.5% |
| This work | MNIST | 28x28 pixels | 28x28→10 | =7,840 | 93.5% |
| | MNIST | 28x28 pixels | 28x28→100→10 | =78,400 | 95.5% |
| | Fashion MNIST | 28x28 pixels | 28x28→100→10 | =78,400 | 91.8% |
| | 26 level EMNIST | 28x28 pixels | 28x28→500→26 | =405,000 | 93.1% |

## 7. MNIST digit classification for multiple channels

To benchmark the optical bandwidth of HITOP. We randomly choose 200 MNIST images from the test set and perform inference on them using HITOP at 7 wavelengths with the 1-layer ANN. The wavelengths are selected to match the corresponding channels of our DWDM. The model has an inference accuracy of 94% while HITOP shows a similar performance at 7 wavelengths from 91% to 93.5%. Which is from 96.8% to 99.5% compared to the electronic inference.

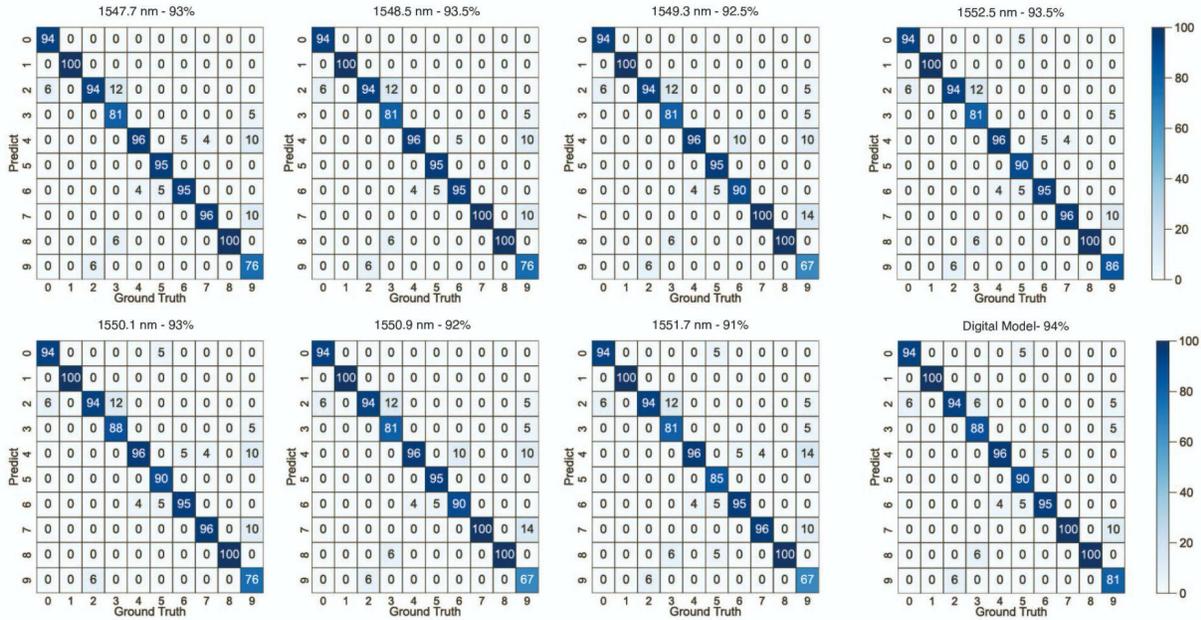

Figs. S10-1 HITOP confusion matrices with 200 MNIST test images at 7 wavelengths.

To verify the parallel computing accuracy of HITOP, we run the MNIST digit inferences with HITOP using 2 wavelength channels simultaneously. 2 sets of 500 MNIST images are randomly picked from the MNIST test dataset. Using the 2-layer DNN model, the electronic processor yields a 96.4% inference accuracy for the first set and 94% for the second set. Firstly, we sent to 2 VCSELs operating at 1549.3 nm and 1550.1 nm, which are the adjacent channels of the DWDM. HITOP shows 95% classification accuracy for the first set and 92.6% for the second set, which is 98.4% and 98.5% compared to the electronic inference.

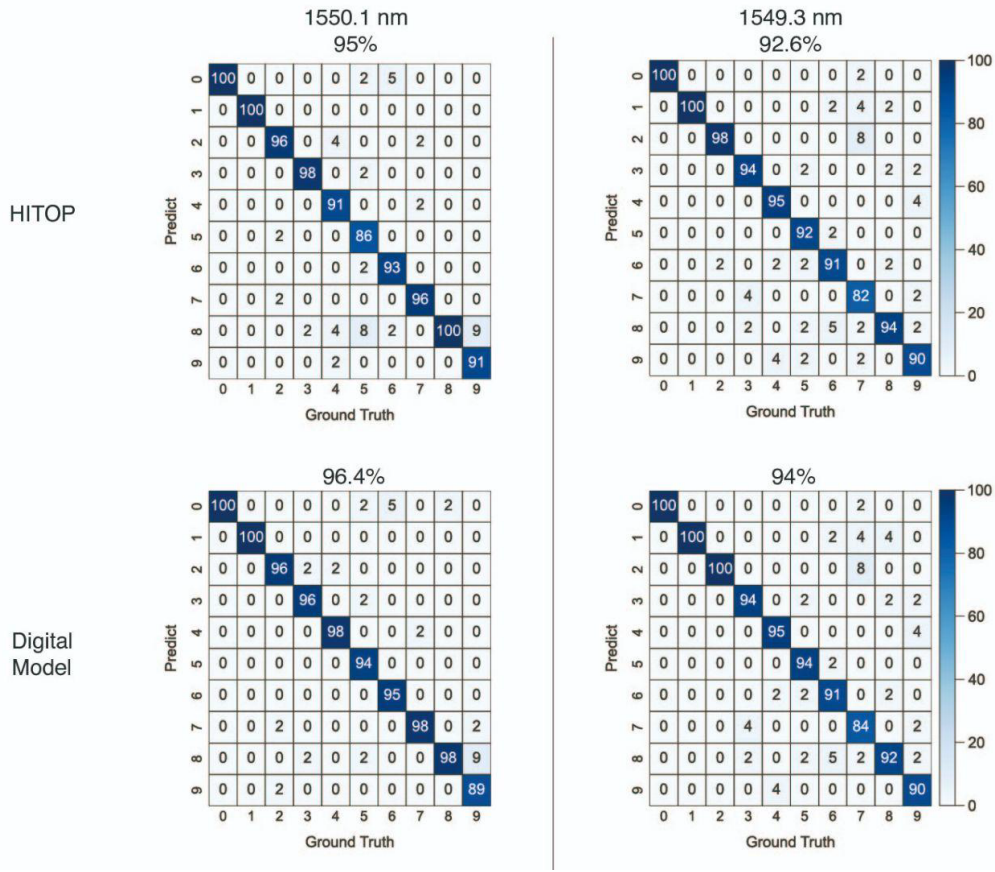

Figs. S10-2 HITOP confusion matrices with 200 MNIST test images at 2 adjacent wavelengths.

We also perform the same inference using 2 further channels in our system at 1548.5nm and 1552.5nm. HITOP shows similar performance with 96% inference accuracy for the first set and 93.6% for the second set, which is 99.6% and 99.6% compared to the electronic inference.

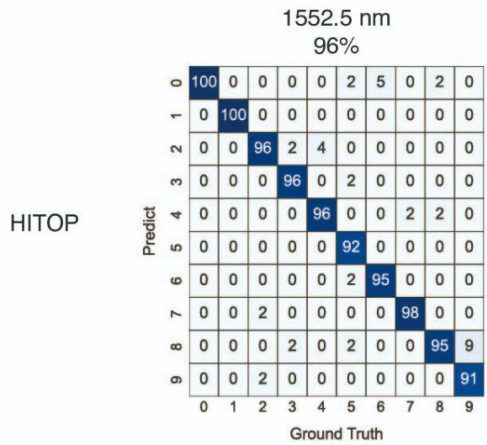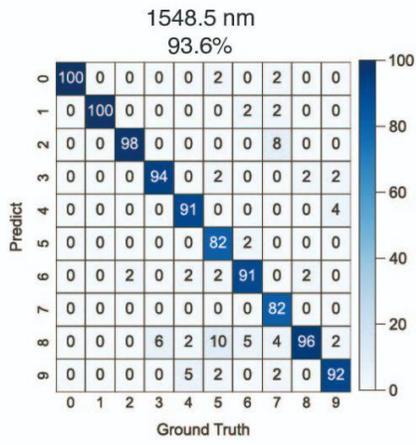
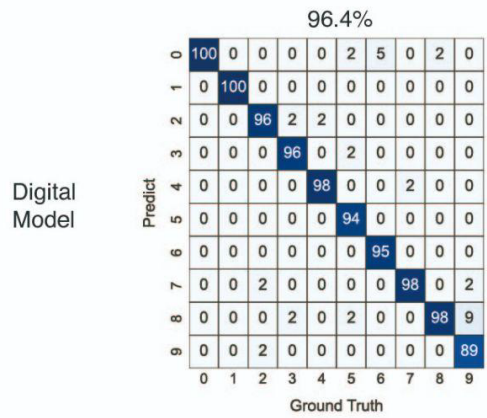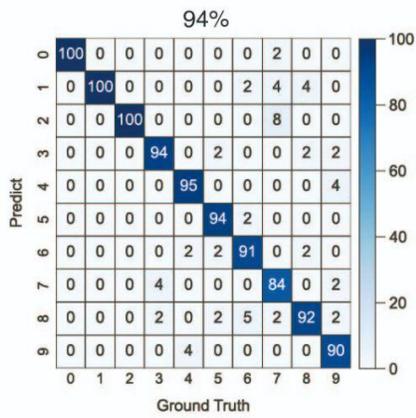

Figs. S10-3 HITOP confusion matrices with 200 MNIST test images at 2 further wavelengths.

All these benchmark tests verify that HITOP has the parallel computing ability and could have a large data throughput when operating at more wavelengths without a large deviation of the computing accuracy.

## 6. Wideband modulation of TFLN modulators

Integrated electro-optic modulator based on thin film lithium niobate can support high speed modulation across a wide optical bandwidth. In figure S11, we show the simulated 3-dB EO bandwidth from 1300 nm to 1800 nm based on the extracted microwave and optical properties of the TFLN EO modulator demonstrated in this paper. Due to the fact that the optical group index only shows small variation across the wide optical bandwidth, the TFLN EO modulation can support > 42 GHz EO modulation bandwidth across 1300-1800nm with a small variation of 0.3 GHz.

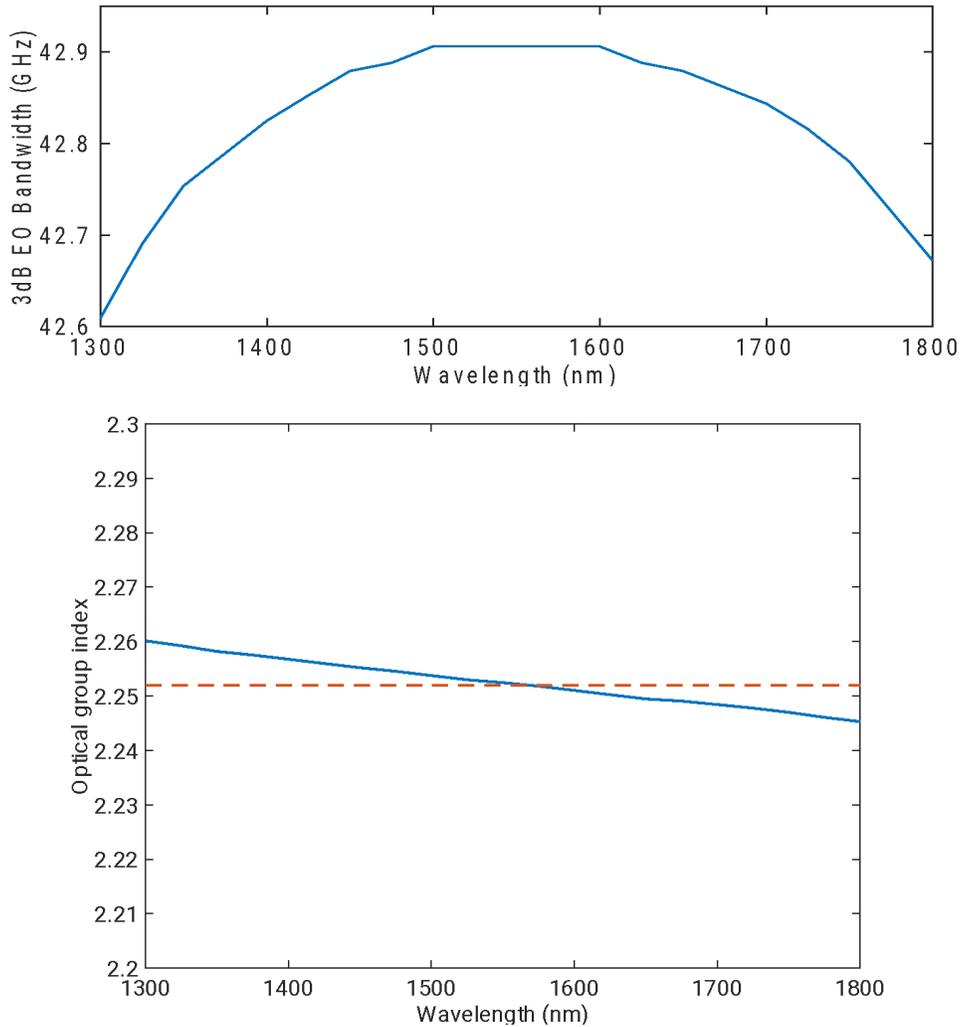

Figure S11 The simulated 3-dB EO bandwidth (top) and optical group index (bottom) as a function of optical wavelength. The EO response simulation is based on the device parameters of the 2-cm-long TFLN modulator device used in this paper, including measured microwave property of the electrodes with 0.7 dB/cm/GHz$^{0.5}$ loss and microwave index of 2.252 (dashed red line). Due to the small variation of the optical group index across the 500-nm optical bandwidth, the 3-dB EO response bandwidth variation is less than 0.3 GHz (from 42.6 GHz at 1300nm to 42.9 GHz at 1550nm).

We further simulation the DC half-wave voltage based on the electrical and optical mode profile [50] at different optical wavelength, shown in Fig 12. Due to the similar EO response curve, both the half-wave voltages at DC and 30 GHz microwave frequency are dominated by the wavevector difference at different optical wavelength. We acknowledge that the difference of the half-wave voltages needs to be pre-calibrated for each VCSEL wavelength.

Finally we provided the measurement results of the same EO modulator at 1.05 μm in Fig ?, which further supports our argument and simulation result that the TFLN modulator can support high speed electro optic modulation across more than 500 nm optical wavelength range. In our paper, 1000 channels each spaced by 50 GHz centered at 1550 nm corresponds to a span from 1375 nm to 1785 nm.

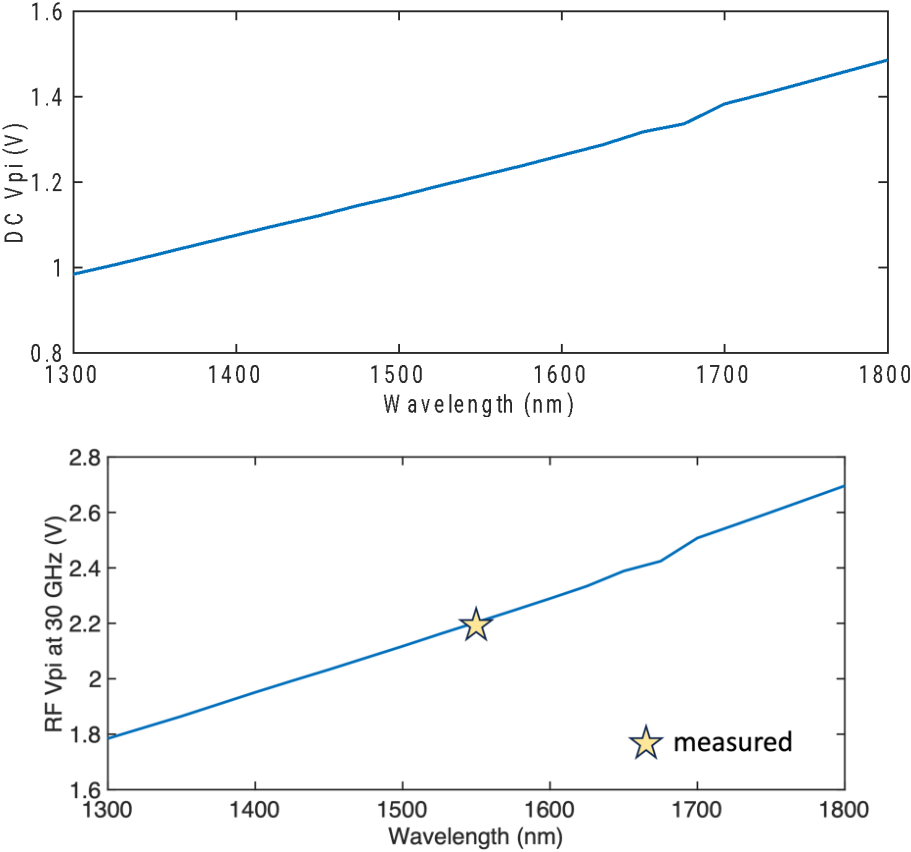

Figure S12 The half-wave voltage at DC (top) and at 30 GHz microwave frequency (bottom) as a function of the optical wavelength.

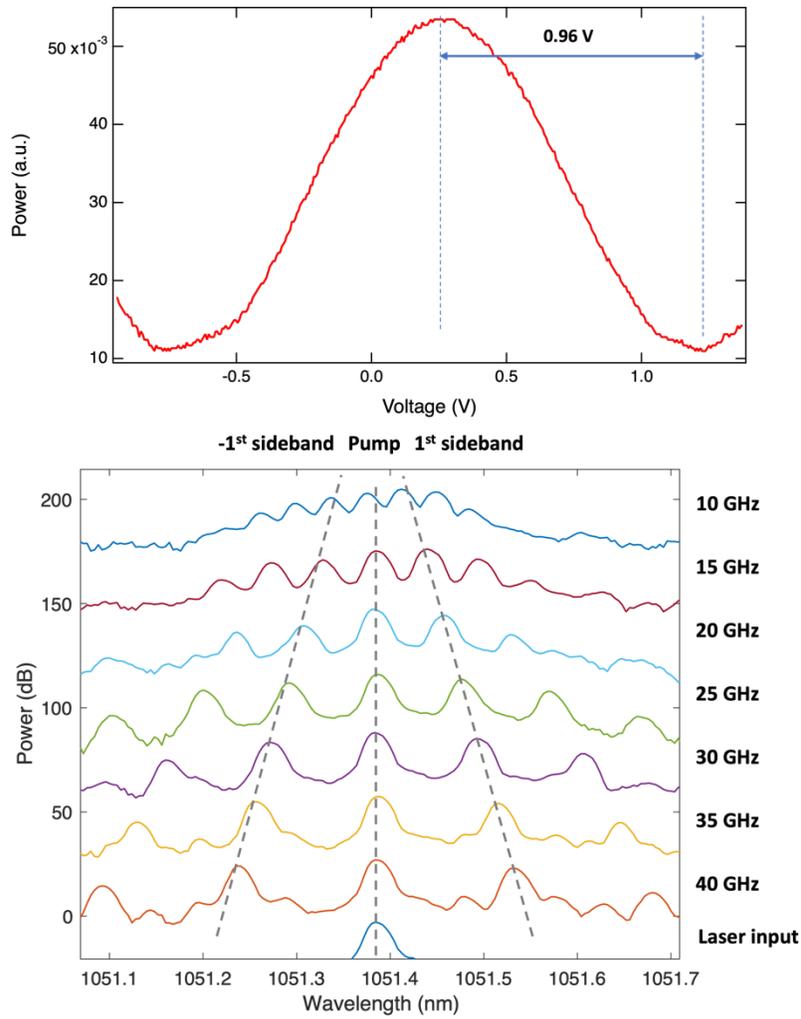

Figure S13 (a) The measured half-wave voltage of 0.96 V at near DC frequency (1MHz) at the optical wavelength of 1.051 μm, as compared to the half-wave voltage of 1.2 V at 1.55 μm. (b) the optical spectrum via driving different microwave frequencies at the optical wavelength near 1 μm. We can extract the half-wave voltage at 30 GHz to be 1.96V, as compared to 2.2 V at 1.55 μm. Both measurement matches well with our simulation results, and it shows the TFLN EO modulator can support a wide range of the optical operating wavelength.